\documentclass[authoryear,preprint,review,12pt]{elsarticle}

\usepackage{graphics}
\usepackage{graphics}

\usepackage{float}
\usepackage{graphicx}
\usepackage{wrapfig}

\usepackage[draft]{todonotes}   

\usepackage[labelfont=bf]{caption}
\usepackage{subcaption}
\usepackage{color}
\usepackage{amsmath} 
\usepackage{mathtools} 
\usepackage{amsfonts} 
\usepackage{amssymb}

\usepackage{array}
\usepackage{booktabs}
\usepackage{rotating}
\usepackage{multirow}
\usepackage{multicol}
\usepackage{lscape}
\usepackage{url}
\usepackage{xcolor}
\usepackage{xargs}                      
\usepackage{calc}

\usepackage{url}

\newcommandx{\MERT}[2][1=]{\todo[linecolor=blue,backgroundcolor=blue!25,bordercolor=blue,#1]{#2}}

\newcommandx{\ISMAIL}[2][1=]{\todo[linecolor=green,backgroundcolor=green!25,bordercolor=green,#1]{#2}}

\usepackage[export]{adjustbox}

\usepackage{algorithm,algorithmicx,algpseudocode}

\usepackage{color}
\usepackage{colortbl}

\usepackage[draft]{todonotes}   
\usepackage{soul}

\definecolor{babyblue}{rgb}{0.54, 0.81, 0.94}
\definecolor{armygreen}{rgb}{0.29, 0.33, 0.13}
\definecolor{brightlavender}{rgb}{0.75, 0.58, 0.89}
\definecolor{aqua}{rgb}{0.0, 1.0, 1.0}
\definecolor{caribbeangreen}{rgb}{0.0, 0.8, 0.6}
\definecolor{reddish}{rgb}{0.82, 0.1, 0.26}
\definecolor{caribbeangreen}{rgb}{0.31, 0.78, 0.47}
\definecolor{jasper}{rgb}{0.84, 0.23, 0.24}
\definecolor{red}{rgb}{1.0, 0.0, 0.0}
\definecolor{green}{rgb}{0.0, 1.0, 0.0}
\definecolor{blue}{rgb}{0.0, 0.0, 1.0}
\definecolor{darkgreen}{rgb}{0.1, 0.7, 0.1}
\definecolor{darkblue}{rgb}{0.1, 0.1, 0.7}
\definecolor{red}{rgb}{0.7, 0.1, 0.1}
\definecolor{coral}{rgb}{1.0, 0.5, 0.31}

\usepackage[colorlinks=true,linkcolor=coral,citecolor=coral]{hyperref}

\newcommand{\squeezeup}{\vspace{-2.5mm}}

\definecolor{darkgreen}{rgb}{0.53, 0.66, 0.42}

\begin{document}

\begin{frontmatter}




\title{Deep Hypergraph U-Net for Brain Graph Embedding and Classification}

\author{Mert Lostar  \fnref{BASIRA}}

\author{Islem Rekik \corref{cor} \fnref{BASIRA,DUNDEE}}
\cortext[cor]{Corresponding author; Dr Islem Rekik (irekik@itu.edu.tr), \url{http://basira-lab.com/}, GitHub code: \url{https://github.com/basiralab/HUNet}}

\author{and for the Alzheimer's Disease Neuroimaging Initiative \corref{ADNI}}
\cortext[ADNI]{Data used in preparation of this article were obtained from the Alzheimer's Disease
Neuroimaging Initiative (ADNI) database (adni.loni.usc.edu). As such, the investigators
within the ADNI contributed to the design and implementation of ADNI and/or provided data
but did not participate in analysis or writing of this report. A complete listing of ADNI
investigators can be found at:
\url{http://adni.loni.usc.edu/wp-content/uploads/how_to_apply/ADNI_Acknowledgement_List.pdf}}

\address[BASIRA]{BASIRA lab, Faculty of Computer and Informatics, Istanbul Technical University, Istanbul, Turkey}
\address[DUNDEE]{School of Science and Engineering, Computing, University of Dundee, UK}




\begin{abstract}
  	
\textbf{-Background.} Network neuroscience examines the brain as a complex system represented by a network (or connectome), providing deeper insights into the brain morphology and function, allowing the identification of atypical brain connectivity alterations, which can be used as diagnostic markers of neurological disorders. 

\textbf{-Existing Methods.} Graph embedding methods which map data samples (e.g., brain networks) into a low dimensional space have been widely used to explore the relationship between samples for classification or prediction tasks. However, the majority of these works are based on modeling the pair-wise relationships between samples, failing to capture their higher-order relationships.

\textbf{-New Method.} In this paper, inspired by the nascent field of geometric deep learning, we propose Hypergraph U-Net (HUNet), a novel data embedding framework leveraging the hypergraph structure to learn low-dimensional embeddings of data samples while capturing their high-order relationships. Specifically, we generalize the U-Net architecture, naturally operating on graphs, to hypergraphs by improving local feature aggregation and preserving the high-order relationships present in the data.

\textbf{-Results.} We tested our method on small-scale and large-scale heterogeneous brain connectomic datasets including morphological and functional brain networks of autistic and demented patients, respectively.

\textbf{-Conclusion.}  Our HUNet outperformed state-of-the art geometric graph and hypergraph data embedding techniques with a gain of 4-14\% in classification accuracy, demonstrating both scalability and generalizability.

\end{abstract}

\begin{keyword}
	Neurological disorder diagnosis, Machine Learning, Computer-Aided Diagnosis, Geometric Deep Learning, Hypergraph UNet
\end{keyword}

\end{frontmatter}

\section{Introduction}

Studying the connectivity of the human brain provides us with a deep understanding of how the brain operates as a  highly complex interconnected system. Network neuroscience, in particular, aims to chart the brain connectome by modeling it as a network, where each node represents a specific anatomical region of interest (ROI) and the weight of the edge connecting pairs of nodes encodes their relationship in function, structure or morphology \citep{fornito:2015,sporns:2019}. Studies of brain networks primarily investigated structural and functional connectivities derived from diffusion weighted and functional magnetic resonance imaging (MRI), respectively \citep{park:2013,sporns:2019}. On a methodological level, graph theory techniques have been widely used to analyze brain networks, giving new insights into atypical alterations of brain connectivity caused by neurological or neuropsychiatric disorders \citep{fornito:2015}. Studies combining these techniques have uncovered that diseases such as schizophrenia \citep{fornito:2012,bloch:2012}, Alzheimer's Disease (AD) \citep{buckner:2009,mahjoub:2018}, autism spectrum disorder (ASD) \citep{morris:2017,soussia:2017} affect the connectomics of the brain, implying that pinning down connectional changes in the brain could reveal clinically useful diagnostic markers.

To this aim, investigating a population of brain connectomes using graph-based embedding techniques has become popular, given their capacity to model the \emph{one-to-one} relationship between data samples (i.e. connectomes) and circumvent the curse of dimensionality in learning tasks such as brain connectome classification or generation. Existing graph embedding techniques can be broken down into three main categories: (1) matrix factorization based, (2) deep learning methods based on random walks and (3) neural network based methods. Matrix factorization focuses on factorizing a high dimensional data matrix into lower dimensional matrices while  preserving the topological properties of the data to factorize. Such methods first encode relationships between nodes into an affinity matrix, which is then factorized to generate the embedding. These vary depending on the properties of the matrix. For instance, while graph factorization (GF) technique uses the adjacency matrix \citep{ahmed:2013}, GraRep \citep{cao:2015} uses $k$-step transition probability matrices. However, matrix factorization based methods usually consider the first order proximity and some of these methods which consider high-order proximities such as the GraRep suffer from scalability issues \citep{goyal:2018}.

Unlike matrix factorization methods, random-walk based deep learning approaches such as DeepWalk \citep{perozzi:2014} and node2vec \citep{grover:2016} focus on optimizing embeddings to encode the statistics of random walks rather than trying to come up with a deterministic node similarity measure. These approaches use random walks in graphs to generate node sequences in order to learn node representations. Given a starting node in a graph, these methods select one of the neighbors and then repeat the process after moving onto the neighboring node to generate node sequences. Random walks have had different uses in approximating different properties in graphs including node similarity \citep{fouss:2007} and centrality \citep{newman:2005}. They are especially helpful when a graph is too large to consider in its entirety or when a graph is only partially observable \citep{goyal:2018}. DeepWalk \citep{perozzi:2014}, one of the initial works using this approach, performs truncated random walks graphs. Instead, node2vec \citep{grover:2016} uses a biased random walk procedure which uses depth first sampling and breadth first sampling together. However, since these approaches use local windows to operate they fail to characterize the global structure of the graph \citep{cai:2018}. 

More recently, there has been a surge of interest in deep graph neural networks (GNN) \citep{kipf:2017,gao:2019,wang:2016}, given their remarkable capacity to model the \emph{deeply nonlinear} relationship between data samples (i.e. connectomes) rooted in message passing, aggregation, and composition rules between node connectomic features \citep{ktena:2017,bessadok:2019,hada:2019,ace:2019}. Graph embedding also witnessed the introduction of different neural networks such as autoencoders \citep{wang:2016}, the multilayer perceptron \citep{tang:2015}, graph convolutional network (GCN) \citep{kipf:2017} and generative adversarial network (GAN) \citep{wang:2017graphgan}. For example, structural deep network embedding (SDNE) \citep{wang:2016}  leveraged deep autoencoders to conserve the information from the first and second order proximities by jointly optimizing both  proximities. Their method applies highly non-linear functions in order to create an embedding that captures the non-linearity of the graph. However this approach can be computationally expensive to operate on large sparse graphs \citep{goyal:2018}. GCN handles this issue by defining a convolution operation for graphs with the aim of iteratively aggregating the embedding neighbors of nodes to update the embedding. Considering only the local neighborhood makes the method scalable while multiple iterations allow for the characterization of the global features.  Graph U-Net (GUNet) \citep{gao:2019} is an encoder-decoder architecture leveraging graph convolution and it improves on GCN by generalizing the seminal U-Net \citep{unet:2015} designed for Euclidean spaces (e.g., images) to non-Euclidean spaces (e.g., graphs), allowing high-level feature encoding and receptive field enlargement through the sampling of important nodes in the graph. 

\begin{wrapfigure}{r}{0.5\textwidth}
  \begin{center}
    \includegraphics[width=0.5\textwidth]{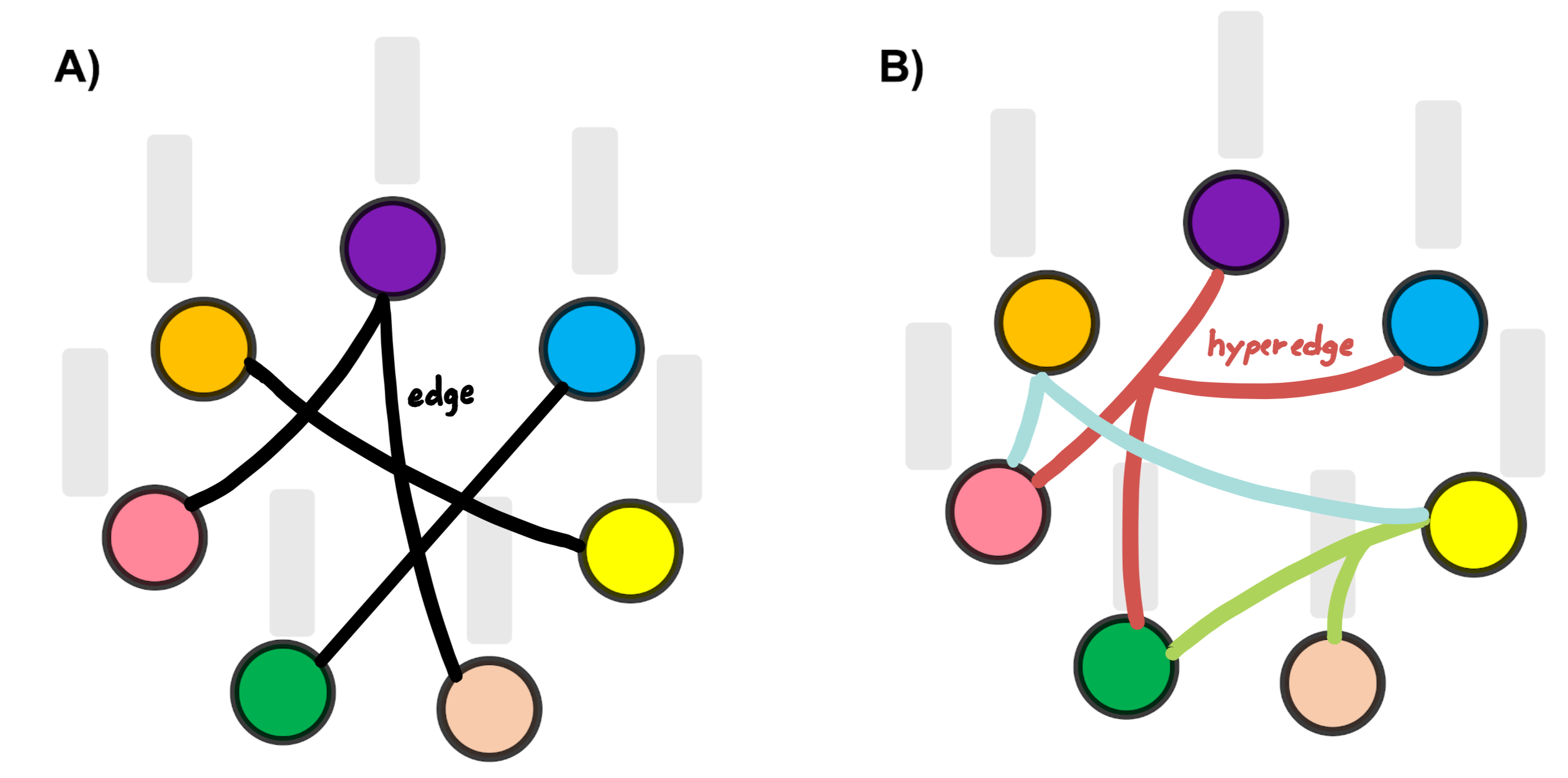}
  \end{center}
  \caption{\textbf{A)} A simple undirected unweighted graph, where an edge connects a pair of nodes. \textbf{B)} An unweighted hypergraph, where a hyperedge connects more than two nodes. While an edge capture the low-order interaction between graph nodes, a hyperedge capture a high-order between nodes as a set. Hence, the learned node embeddings (gray vertical bars) in a hypergraph better capture complex and representative node interactions.}
  \label{wrapfig:1} 
\end{wrapfigure}

More interestingly, such graph embedding architectures allow to circumvent the curse of dimensionality in learning based tasks such as brain connectome classification or generation \citep{bessadok:2019,hada:2019} by learning low-dimensional embeddings of node attributes such as connectome features while preserving their similarities. These learned embeddings can then be used as inputs for machine learning methods for tasks such as node classification \citep{hada:2019,ace:2019} and link prediction \citep{liu:2017}. However, a major limitation of current deep graph embedding architectures is that they are unable to capture \emph{many-to-many} (i.e., high-order) relationships between samples, hence the learned feature embeddings only consider node-to-node edges in the population graph.

Hypergraph Neural Network (HGNN) \citep{feng:2019} addresses this problem through the use of hypergraph structure for data modeling. The main difference between a traditional graph and a hypergraph, as illustrated in (\textbf{Fig.}~\ref{wrapfig:1}), is that graphs are only able to represent one-to-one node relationships via edges while hypergraphs are able to capture high-order relationships between nodes via the concept of a \emph{hyperedge connecting a subset of nodes}. Even tough the traditional hypergraph learning approach usually suffers from high computational costs, HGNN manages to eliminate this challenge by devising a hyperedge convolution operation. However, HGNN only uses the devised hypergraph convolution operation for learning  the hypernode  embeddings. 

In this paper we propose the Hypergraph U-Net (HUNet) architecture for high-order data embedding by generalizing the graph U-Net \citep{gao:2019} to hypergraphs. HUNet, unlike HGNN takes advantage of the U-Net architecture to improve the local feature aggregation through pooling and unpooling operations while still leveraging the hypergraph convolution operation to learn more representative feature embeddings. It enables the inferring and aggregation of node embeddings while exploring the \emph{global high-order} structure present between \emph{subsets} of data samples, becoming more gnostic of real data complexity.

We evaluate HUNet on two different types of brain connectomic datasets for neurological disorder diagnosis and show that HUNet achieves a large gain in classification accuracy compared to state-of-the-art methods on both datasets, demonstrating scalability and generalizability as we perturb training and test sets and vary their sizes.

\begin{figure}[H]
\centering
\includegraphics[width=14cm]{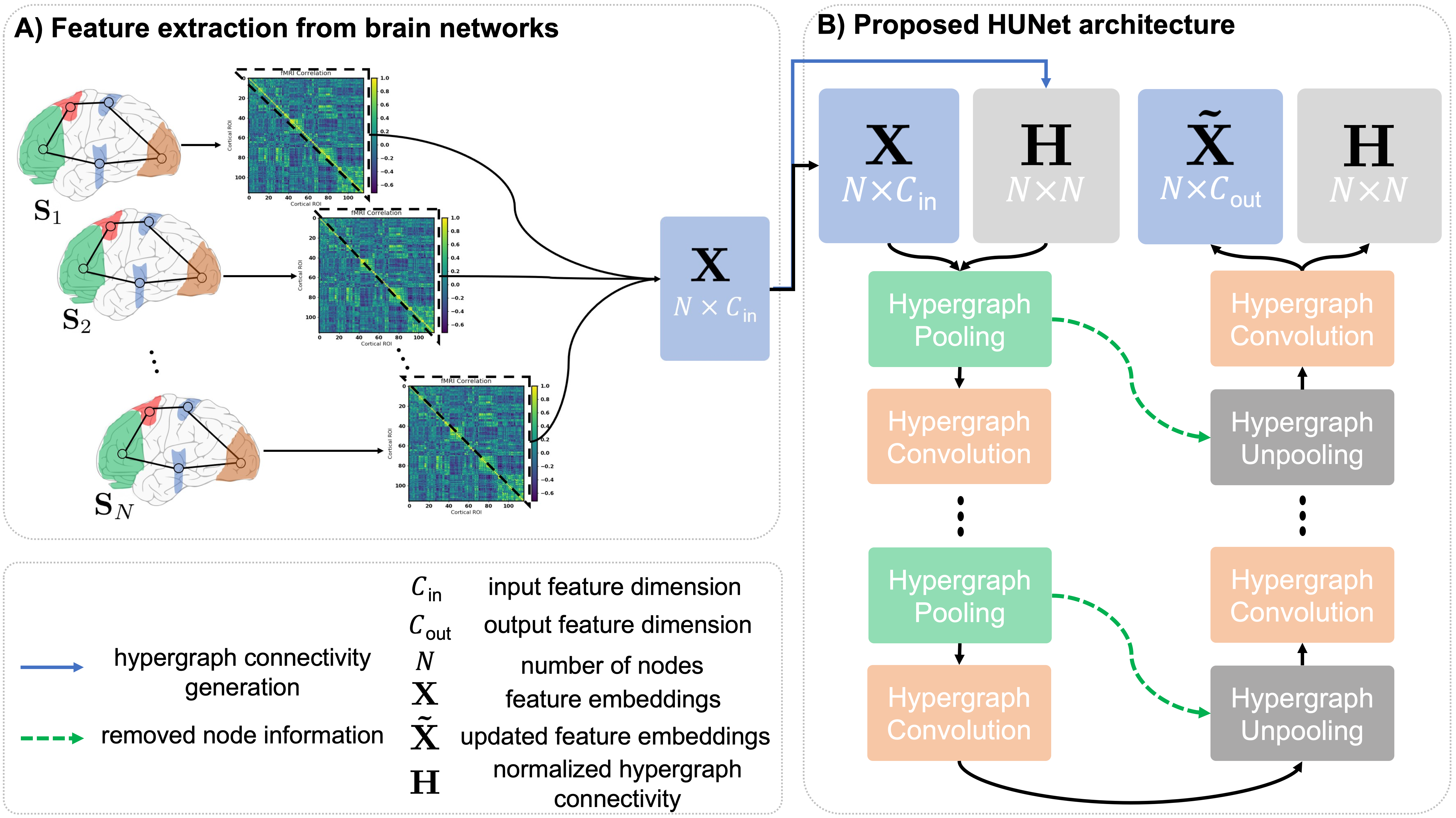}
\caption{\emph{Hypergraph U-Net (HUNet) architecture.} \textbf{A)} Each brain network $\mathbf{S}_i$ of subject $i$ is encoded in a connectivity matrix, which is vectorized by extracting $C_{in}$ connectivity weights stored in its off-diagonal upper triangular part. We stack all $N$ samples into $\mathbf{X}$ with $C_{in}$ rows. \textbf{B)} Using $\mathbf{X}$, we generate the normalized hypergraph connectivity matrix $\mathbf{H} \in \mathbb{R}^{N \times N}$. $\mathbf{H}$ and $\mathbf{X}$ are used as inputs for the proposed HUNet architecture, stacking our proposed hypergraph pooling (hPool) and unpooling (hUnpool) layers with hypergraph convolution layers. Connectivity information of the removed nodes at hPool layer at HUNet level $d$ are transferred to the hUnpool layer at the same level to be used when up-sampling $\mathbf{X}$ and restoring $\mathbf{H}$. Outputs of the HUNet are the learned feature embeddings $\tilde{\mathbf{X}}$, which can be used for the target learning task.}  
\label{fig:4} 

\end{figure}

\begin{table}[thp]
\caption{ \emph{Major mathematical notations used in this paper}.\label{table:1}}
\centering
\begin{scriptsize}
\begin{tabular}{c@{~~}c@{~~}c}
\toprule
	Mathematical notation & Definition \\
	\midrule
	$\mathbf{S}_{i}$ & brain network of subject $i$\\ 
	$N$ & number of nodes in the initial hypergraph\\
	$\mathbf{X} \in \mathbb{R}^{N \times C_{in}}$ & input feature embeddings where $C_{in}$ is the input feature dimension \\
	$E$ & number of hyperedges in the initial hypergraph\\
	$N_{d}$ & number of nodes at HUNet level $d$\\
	$\mathbf{X}_{d} \in \mathbb{R}^{N_{d} \times C_{d}}$ & feature embeddings at HUNet level $d$, $C_{d}$ is the feature dimension \\
	
$\mathbf{Q} \in \{0,1\}^{N \times E} $ & hypergraph incidence matrix\\
$\mathbf{Q}(v,e) = \begin{cases}
           0, if v \notin e\\
           1, if v \in e
        \end{cases}$ & $\mathbf{Q}$ entries where $v$ is a vertex and $e$ is a hyperedge \\

$d(v) =  \sum_{e \in \mathcal{E}}\mathbf{Q}(v,e)$ & degree of a vertex\\
$d(e) = \sum_{v \in V} \mathbf{Q}(v,e) $ & degree of a hyperedge \\
$\mathbf{D}_{v} \in \mathbb{R}^{N \times N}$ & diagonal matrix of vertex degrees \\
$\mathbf{D}_{\varepsilon} \in \mathbb{R}^{E \times E}$ & diagonal matrix of hyperedge degrees \\
$\mathbf{W} \in \mathbb{R}^{E \times E}$ & diagonal weight matrix of a hypergraph\\

 $ \mathbf{H} = {\mathbf{D}_{v}}^{\frac{-1}{2}} \mathbf{Q}\mathbf{W}{\mathbf{D}_{\varepsilon}}^{-1} \mathbf{Q}^{T}{\mathbf{D}_{v}}^{\frac{-1}{2}} \in \mathbb{R}^{N \times N}$ & normalized hypergraph connectivity based on $\mathbf{Q}$ incidence matrix \\
 
 $\mathbf{\Theta}_d \in \mathbb{R}^{C_{d} \times C_{d+1}}$& a learnable matrix where $C_{d}$ is the input feature size \\ & and $C_{d+1}$ is the output feature size\\
 
 $\mathbf{Z}_{d} \in \mathbb{R}^{N_{d} \times C_{d+1}}$  & updated feature embeddings at HUNet level $d$\\
 
 $ \mathbf{Z}_{d} = \mathbf{H}_{d}\mathbf{X}_{d}\mathbf{\Theta}_{d} $ & hypergraph convolution operation at HUNet level $d$\\

 \midrule
\end{tabular}
\end{scriptsize} 
\end{table}
\section{Proposed method}

\subsection{Previous Works}
In graph embedding the aim is learning how to project node features into a low-dimensional space while preserving the structural relationships in the graph such that nodes with links are also close to each other in this new embedding space. These node embeddings can then be used as inputs for machine learning methods to tackle a variety of prediction and network analysis tasks. Our work builds upon recent advances in graph embedding and hypergraph learning techniques.\\

\subsubsection{Graph U-Net embedding architecture}
A wide variety of graph embedding models using different approaches have been proposed and applied to tasks such as node classification \citep{perozzi:2014,tang:2015}, node clustering \citep{tang:2016}, link prediction \citep{wang:2016}, graph alignment \citep{hada:2019}, graph classification \citep{dai:2016,ace:2019} and graph visualization \citep{cao:2016} in the recent years. More recently, Graph U-Net (GUNet) \citep{gao:2019} was proposed as an U-Net like architecture for graph data. By adapting Euclidean pooling and unpooling operations, which are critical when building encoder-decoder architectures, to non-Euclidean graph data with no spatial locality and order information. GUNet comprises Graph Convolutional Networks (GCN) layers \citep{kipf:2017} to learn deeply composed embeddings of the graph nodes by exploring their hierarchical topological neighbors via the `neighbor of a neighbor' composition rule. They show that this encoder-decoder architecture outperforms conventional GCN in learning well representative embeddings of the node features. However, GUNet is only able to learn from \emph{pair-wise relationships} between the nodes, thereby ignoring the \emph{many-to-many} high-order relationships present in many real-world data.

\subsubsection{Hypergraph learning}
Very recently, hypergraphs, originally introduced in \citep{zhou:2006}, have started to gain momentum in geometric deep learning thanks to their ability to capture high-order relationship between data samples in various tasks such as feature selection \citep{zhang:2017}, image classification \citep{yu:2012}, social network analysis \citep{fang:2014} and sentiment prediction using multi-modal data \citep{ji:2018}. More recently, the inception of hypergraph neural network (HGNN) \citep{feng:2019} as the first geometric deep learning model on hypergraph structure introduced hyperedge convolution operations based on the hypergraph Laplacian encoding the hypergraph spectra. However, HGNN is restricted to using hypergraph convolution for learning hypernode embeddings in a semi-supervised manner where training node labels supervise the estimation of the node feature mapping from layer to layer. In the absence of node labels, HGNN cannot be trained. 
Besides, HUNet takes advantage of the U-Net architecture, aiming to learn well representative feature embeddings by improving the first-order local feature aggregation through \emph{pooling and unpooling} operations while combining them with hypergraph convolution operation for improving global feature aggregation compared to applying \emph{convolution alone} for feature aggregation as in HGNN architecture \citep{feng:2019}. 

\subsection{Proposed Geometric Hypergraph U-Net (HUNet)}
In this section we explain our proposed HUNet architecture, shown in (\textbf{Fig.}~\ref{fig:4}) and its components. The key idea behind the HUNet architecture is to learn a \emph{many-to-many} node embedding with a high-order feature aggregation rule by leveraging the advantage of using hypergraphs to model the high-order relations between hypernodes compared to existing deep graph-based embedding methods. With this purpose, we propose the hypergraph pooling (hPool) and hypergraph unpooling (hUnpool) layers.

\subsubsection{Hypergraph convolution}
Even tough graphs are adequate for representing pair-wise relationships between different nodes, in many applications higher-order relationships, which graphs are unable to represent, are present between subsets of nodes. For such applications one can take advantage of the hypergraph structure encoding shared interactions between a subset of nodes by connecting them with a single hyperedge \citep{zhou:2006} (\textbf{Fig.}\ref{wrapfig:1}). We define the basic hypergraph as $G = \{V, \mathcal{E}, \mathbf{W}\}$, where $V$ is a node set, $\mathcal{E}$ is a hyperedge set, $\mathbf{W} \in \mathbb{R}^{E \times E}$ is a diagonal weight matrix, where $E$ is the number of hyperedges, assigning weights to hyperedges. In our experiments $\mathbf{W}$ is initialized as an identity matrix meaning that all the hyperedges have the same weight. We then define a $N \times E$ hypergraph incidence matrix $\mathbf{Q}$, where $N$ is the number of hypernodes in the hypergraph, with elements representing whether a hypernode is contained in a hyperedge or not as follows:

\begin{equation}
    \mathbf{Q}(v,e) = \begin{cases}
               0, if v \notin e\\
               1, if v \in e
            \end{cases}
\end{equation}

where $v \in V$ represents a hypernode and $e \in \mathcal{E}$ represents a hyperedge. We also define the hyperedge degree $d(e)$ which represents the number of hypernodes in a hyperedge $e$ and node degree $d(v)$ denoting the number of hyperedges connected to a hypernode $v$ as:

\begin{align}
    d(e) &= \sum_{\mathclap{v \in V}} \mathbf{Q}(v,e)
 &
    d(v) &= \sum_{\mathclap{e \in \mathcal{E}}} \mathbf{Q}(v,e)
\end{align}

In order to adapt the hypergraph convolution operation to our encoder-decoder U-Net architecture, given a hypergraph with $N$ hypernodes, we produce the normalized hypergraph connectivity $\mathbf{H} \in \mathbb{R}^{N \times N}$ as follows:

\begin{equation}
    \mathbf{H} = {\mathbf{D}_{v}}^{\frac{-1}{2}}\mathbf{Q}\mathbf{W}{\mathbf{D}_{\varepsilon}}^{-1}\mathbf{Q}^{T}{\mathbf{D}_{v}}^{\frac{-1}{2}}   
\end{equation}

where $\mathbf{D}_{v} \in \mathbb{N}^{N \times N}$ denotes the diagonal hypernode degree matrix and $\mathbf{D}_{\varepsilon} \in \mathbb{N}^{E \times E}$ represents the diagonal hyperedge degree matrix.  $\mathbf{H}$, constructed from the initial hyperpraph, is also pooled and unpooled along with the feature embeddings in the pooling and unpooling layers but unlike the feature embeddings, it only changes in dimensionality. Normalized hypergraph connectivity  $\mathbf{H}_{d} \in \mathbb{R}^{N_{d} \times N_{d}}$ at HUNet level $d$, where $N_{d}$ is the number of nodes in the hypergraph at level $d$, is pooled and restored by the respective pooling and unpooling layers at level $d$. The feature embedding $\mathbf{X}_{d} \in \mathbb{R}^{N_{d} \times C_{d}}$, where $C_{d}$ is the feature dimension at level $d$, is taken as input from the previous pooling or unpooling layer along with $\mathbf{H}_{d}$. The hypergraph feature matrix $\mathbf{X}_{d}$ is \emph{first transformed} by the embedding function $\mathbf{\Theta}_d \in \mathbb{R}^{C_{d} \times C_{d+1}}$ learned by the hypergraph convolution layer at level $d$ to extract $C_{d+1}$ dimensional node features, \emph{then diffused} through $\mathbf{H}_{d}$ to aggregate the embedded hypernode features across hyperedges that contain them. The hypernode feature embedding matrix $\mathbf{Z} \in \mathbb{R}^{N_{d} \times C_{d+1}}$ is then passed on to the next HUNet level and updated as follows:

\begin{equation}
\label{eq:hyper}
\mathbf{Z}_d = \mathbf{H}_{d}\mathbf{X}_{d}\mathbf{\Theta}_d
\end{equation}

\subsubsection{Hypergraph pooling layer}
\begin{figure}[H]
\centering
\includegraphics[width=14cm]{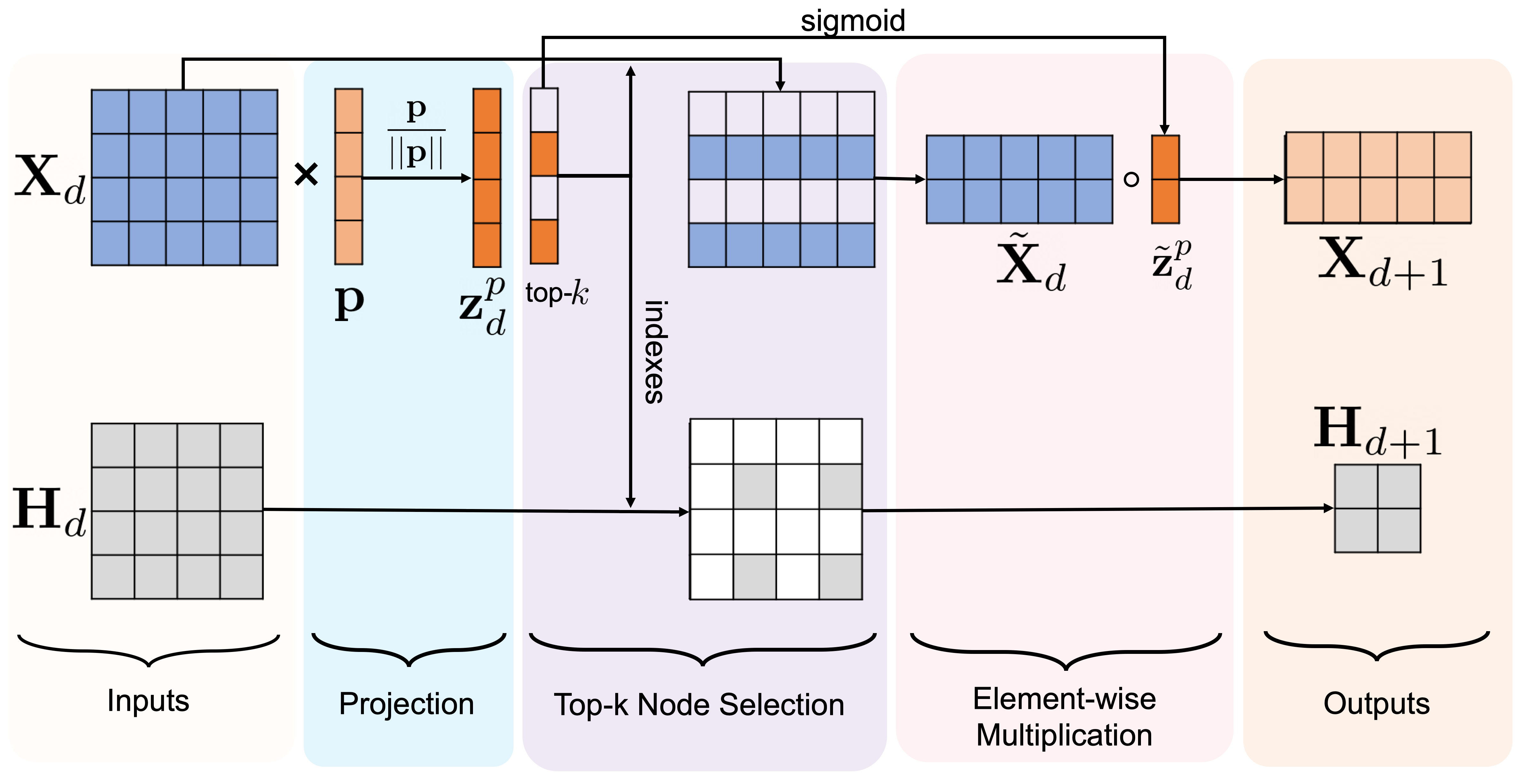}
\caption{\emph{HUNet pooling layer (hPool).} The feature embedding $\mathbf{X}_{d} \in \mathbb{R}^{N_{d} \times C_{d}}$ and hypergraph connectivity matrix $\mathbf{H}_{d} \in \mathbb{R}^{N_{d} \times N_{d}}$  are passed to the hPool layer as inputs where $N_{d}$ is the number of nodes, $C_{d}$ is the feature dimension at HUNet level $d$. In the projection phase, the feature embedding is projected on the learnable vector $\mathbf{p} \in \mathbb{R}^{N_{d}}$. The output of this operation, represented by $\mathbf{z}_{d}^{p} \in \mathbb{R}^{N_{d}}$, is then used to determine the indices of the top-$k$ nodes. This index information is what we use in order to pool our inputs in the top-$k$ node selection phase. This pooling operation produces $\tilde{\mathbf{X}}_{d} \in \mathbb{R}^{k \times {C}_{d}}$ which represents the pooled feature embedding and $\mathbf{H}_{d+1} \in \mathbb{R}^{k \times k}$ denoting  the hypergraph connectivity which we pass onto the next HUNet level.}
\label{fig:2} 
\end{figure}

We propose a hypergraph pooling (hPool) layer to down-sample our hypergraph. Instead of passing the graph adjacency matrix to our pooling layer as in \citep{gao:2019}, we use the $\mathbf{H}$ described in the previous part in order to adapt the pooling operation to hypergraphs (\textbf{Fig.}~\ref{fig:2}). This layer is used to choose a subset of hypernodes that form a smaller hypergraph while losing as little information as possible. In order to learn how to select such hypernodes, a trainable projection vector $\mathbf{p} \in \mathbb{R}^{N_{d}}$ is used to map all hypernode features to a real-valued score. This projection allows the use of top-$k$ hypernode pooling for selecting the $k$ most important hypernodes. The individual hypernode scores represent how much information is preserved after the projection onto the $\mathbf{p}$ vector. This means that selecting the top scoring $k$-hypernodes to form the new hypergraph would maximize information preservation. We define the layer-wise propagation rule for this pooling layer as follows:

\begin{equation}
\begin{aligned}
\mathbf{z}_{d}^{p} &= \mathbf{X}_{d}\frac{\mathbf{p}}{||\mathbf{p}||}\\
indexes &= topk(\mathbf{z}_{d}^{p},k) \\
\tilde{\mathbf{z}}_{d}^{p} &= sigmoid(\mathbf{z}_{d}^{p}[indexes])\\
\tilde{\mathbf{X}}_{d} &= \mathbf{X}_{d}[indexes,:]\\
\mathbf{X}_{d+1} &= \tilde{\mathbf{X}}_{d} \odot \tilde{\mathbf{z}}_{d}^{p}\\
\mathbf{H}_{d+1} &= \mathbf{H}_{d}[indexes,indexes] 
\end{aligned}
\label{eq:pooling}
\end{equation}

\normalsize
where $\mathbf{X}_{d} \in \mathbb{R}^{N_{d} \times C_d}$ and $\mathbf{H}_{d} \in \mathbb{R}^{N_{d} \times N_{d}}$ denote the feature embedding and hypergraph connectivity, respectively, at depth $d$ with $C_d$ denoting the feature embedding dimension. $\mathbf{z}_{d}^{p} \in \mathbb{R}^{N_{d}}$ is output of the projection of $\mathbf{X}_{d}$ onto $\mathbf{p}$. $topk(\mathbf{z}_{d}^{p},k)$ operation returns the indexes of nodes with the largest scores in $\mathbf{z}_{d}^{p}$. These indexes are then used to produce the pooled feature embeddings $\tilde{\mathbf{X}}_{d} \in \mathbb{R}^{k \times C_d}$ and the pooled hypergraph connectivity $\mathbf{H}_{d+1} \in \mathbb{R}^{k \times k}$. We select the largest entries from $\mathbf{z}_{d}^{p}$ and apply a sigmoid function to produce  $\tilde{\mathbf{z}}_{d}^{p}$. Next, we apply an element-wise multiplication, represented by $\odot$ to $\tilde{\mathbf{X}}_{d}$ and $\tilde{\mathbf{z}}_{d}^{p}$, thereby generating the new feature embedding $\mathbf{X}_{d+1}$ to pass onto the next HUNet level $(d+1)$ along with $\mathbf{H}_{d+1}$. Note that only the dimensionality of the hypergraph connectivity changes from layer to layer during both encoding (i.e., pooling) and decoding (i.e., unpooling) steps in the HUNet architecture (\textbf{Fig}~\ref{fig:4}).

\subsubsection{Hypergraph unpooling layer}
An unpooling operation is needed in order to up-sample the hypergraph data that was previously pooled in the hypergraph encoding phase. To this end, we propose hypergraph unpooling layer (hUnpool) that generalizes the unpooling operation proposed in \citep{gao:2019} to hypergraphs by leveraging the normalized hypergraph connectivity $\mathbf{H}$. This layer takes the connectivity information about the removed nodes from the hPool layer at the same level of the HUNet to reconstruct the hypergraph and place back the removed nodes (\textbf{Fig}~\ref{fig:4}). However, the nodes are placed back with empty feature vectors which are filled in using the hypergraph convolution operation (\textbf{Eq.}~\ref{eq:hyper}).

Using a hyper U-Net instead of a graph U-Net gives the ability to preserve high-order sample interactions when mapping data samples into a low-dimensional space. The normalized hypergraph connectivity $\mathbf{H}$ drives the hypergraph convolution in both encoding (top-down) and decoding phases (bottom-up) as illustrated in \textbf{Fig}~\ref{fig:4}. Blocks of hypergraph pooling and hypergraph unpooling layers are stacked in order to construct the HUNet architecture. Hypergraph convolution follows every hypergraph pooling layer to update the features of the nodes using their first-order local neighbors as well as every hypergraph unpooling layer to fill in the empty node features that were added back. The algorithm of our HUNet architecture with  $d$-depth is detailed in \textbf{Algorithm.}~\ref{algo:1}.

\begin{algorithm*}
\caption{A Hypergraph U-Net of depth $d$}\label{algo:1}
\begin{scriptsize} 
\begin{algorithmic}[1]

\State \textbf{Definitions:} $x_{s}$: array of feature embeddings (empty at initialization)\\
    $h_{s}$: array of normalized hypergraph connectivities (empty at initialization)\\
    $N$: number of rows in $\mathbf{X}$\\
    $idx_{s}$: array of indices (empty at initialization)\\
    $idx$: selected node indices in top-$k$ pooling\\
    $hPool(\dot)$: hypergraph pooling layer\\
    $r$: pooling ratio to be used in the hPool layers\\
    $hConv()$: hypergraph convolution\\
    $uact()$: activation function that is used between HUNet levels\\
    $zeros(input)$: creates a zero matrix in the shape of the input matrix\\
    $oact()$: output activation function
\State \textbf{INPUTS:} 
	$\mathbf{X}$: feature embeddings;
	$\mathbf{H}$: hypergraph connectivity;
	    \State$x_{s}.append(\mathbf{X})$
	    \State$h_{s}.append(\mathbf{H})$
	    \State$idx_{s}.append([1,2 \dots N])$
	    \For{depth = 1, 2, \dots, d }
	        \State$\mathbf{X},\mathbf{H},idx = hPool(\mathbf{X},\mathbf{H},r)$
	        \State$\mathbf{X} =  uact(hConv(\mathbf{X},\mathbf{H}))$
	        \State $idx_{s}.append(idx)$
	        \If{$depth<d$}
       \State $x_{s}.append(\mathbf{X})$
	        \State $\mathbf{H}_{s}.append(\mathbf{H})$
    \EndIf
		    \EndFor
	    \For{depth = d - 1, d - 2, \dots, 0}
	        \State $\mathbf{ret\_x} = x_{s}[depth]$
	        \State $\mathbf{ret\_\mathbf{H}} = h_{s}[depth]$
	        \State $ret\_idx = idx_{s}[depth+1]$
	        \State $\mathbf{upsample} = zeros(\mathbf{ret\_x})$
	        \State $\mathbf{upsample}[ret\_idx] = \mathbf{X}[ret\_idx]$
	        \State $\mathbf{X} = \mathbf{ret\_x} + {\mathbf{upsample}}$
	        \State $\mathbf{X} = hConv(\mathbf{X},\mathbf{ret\_\mathbf{H}})$
	        \If{$depth>0$}
       \State $\mathbf{X} = uact(\mathbf{X})$
       \Else
	        \State $\mathbf{X} = oact(\mathbf{X})$
    \EndIf
    \EndFor
	\State \textbf{OUTPUT: } learned hypernode feature embedding $\mathbf{X}$ 

\end{algorithmic}
\end{scriptsize} 
\end{algorithm*}

\section{Results}

\subsection{Evaluation Dataset} 

We evaluate our HUNet and comparison state-of-the-art methods on small-scale and large-scale connectomic datasets derived from different neuroimaging modalities (structural and functional MRI) to demonstrate the ability of HUNet in better generalizing across data scales and handling heterogeneous data distributions.

\textbf{Small-scale morphological data} We use a subset of ADNI GO\footnote{\url{http://adni.loni.usc.edu}} public dataset, consisting of 77 subjects (41 AD and 36 Late Mild Cognitive Impairment), where each subject has a structural T1-w MR image \citep{mueller:2005}.  Data used in the preperation of this article were obtained from the Alzheimer's Disease Neuroimaging Initiative (ADNI) database (adni.loni.usc.edu). The ADNI was launched in 2003 as a public-private partnership, led by Principal Investigator Michael W. Weiner, MD. The 
primary goal of ADNI has been to test whether serial magnetic resonance imaging (MRI), positron emission 
tomography (PET), other biological markers, and clinical and neuropsychological assessment can be combined 
to measure the progression of mild cognitive impairment (MCI) and early Alzheimer's disease (AD).
For preprocessing, we follow the steps defined by \citep{mahjoub:2018}. In order to reconstruct left and right cortical hemispheres from T1-w MRI \citep{fischl:2004}, FreeSurfer \citep{fischl:2012} processing pipeline was used for each subject. Next, each cortical hemisphere was divided into 35 cortical regions using Desikan-Killany cortical atlas \citep{fischl:2004}. We then use cortical attributes: maximum principal curvature, cortical thickness, sulcal depth and average curvature to derive four $35 \times 35$ morphological brain connectivity matrices. For each attribute, we extract a feature vector by retrieving the off-diagonal upper triangular part elements of each attribute-specific connectivity matrix.

\textbf{Large-scale functional network data} We also evaluate our method on a large scale functional network dataset, consisting of 517 subjects (245 ASD and 272 Control) from the ABIDE\footnote{\url{http://preprocessed-connectomes-project.org/abide/}} preprocessed dataset \citep{ABIDE:2013}. Different preprocessing steps were carried out by the
data processing assistant for resting-state fMRI (DPARSF) pipeline, which
is established on statistical parametric maps (SPM) and resting-state fMRI data
analysis toolkit (REST). In order to ensure a steady signal, first 10 volumes
of rs-fMRI images were abandoned. Based on a six-parameter (rigid body), all
images were slice timing corrected and realigned to the middle in order to cut
down on inter-scan head motion \citep{tang:2018}. After this step, the functional
data were registered in Montreal Neurological Institute (MNI) space using a
resolution of $3 \times 3 \times 3$ $mm^{3}$. In order to boost the signal to noise ratio, spatial smoothing was then applied using a Gaussian kernel of 6 mm. Lastly, a band-pass filtering (0.01-0.1 Hz) was applied to the time series of each voxel \citep{price:2014,huang:2017}. Detailed explanations for these steps can be found in \url{http://preprocessed-connectomes-project.org/abide/}. Each brain rfMRI was partitioned into 116 ROIs to construct $116 \times 116$ connectivity matrices where we select the upper off-diagonal triangles as feature vectors for the individual subjects (hypernodes).


\subsection{Evaluation and comparison methods} 
\textbf{Performance measures}
We evaluated the performance of the methods for node classification by using the results derived from the morphological and functional connectomic datasets in terms of accuracy, sensitivity and specificity. For the ADNI dataset we also averaged the results on the four morphological attributes to calculate an overall result.

\textbf{Parameter setting} The initial hypergraph was constructed from the features using $k$-nearest neighbors algorithm with $k = 2$ for the morphological dataset and $k = 4$ for the functional dataset as it has more nodes. We used $depth =2$ for both GUNet and HUNet and $pooling\_ratio = 0.8$ for HUNet and $pooling\_ratio = 0.5$ for GUNet in both datasets. We set the learning rate to $0.01$ across architectures and datasets. For HGNN, we used 2 hypergraph convolutional layers with a dropout layer in-between at $0.5$ rate.

\textbf{Evaluation} \textbf{Table.}~\ref{table:2} compares the performance of HUNet to our baselines on the morphological connectomic dataset in terms of classification accuracy, sensitivity and specificity. Clearly, HUNet outperforms the baseline methods on most connectomic views, improving the classification accuracy by a margin of $\sim$7-14\%. As for the large-scale functional dataset, HUNet outperformed other baselines by $\sim$4\% as shown in  \textbf{Table.}~\ref{table:3}.  The results show that HUNet achieves a classification accuracy gain of both $\sim$3-14\% across multi-scale heterogeneous datasets in comparison with state-of-the-art methods. This shows that our model is scalable and generalizable. 

\begin{table}[ht]
\centering
\tiny
\renewcommand{\tabcolsep}{0.10cm}
\begin{tabular}{l|ccc|ccc|ccc|ccc|ccc}

 \emph{Model} & \multicolumn{3}{c|}{\emph{Max principal curvature}} & \multicolumn{3}{c|}{\emph{Cortical thickness}} & \multicolumn{3}{c|}{\emph{Sulcal depth}} & \multicolumn{3}{c|}{\emph{Average curvature}} & \multicolumn{3}{c}{\emph{Overall}} \\ 
 & {ACC} & {SEN} & {SPEC} & {ACC} & {SEN} & {SPEC} & {ACC} & {SEN} & {SPEC} & {ACC} & {SEN} & {SPEC} & {ACC} & {SEN} & {SPEC}  \\
  \hline
  GUNet \citep{gao:2019}   & $\mathbf{86.7\%}$ & $\mathbf{83\%}$ & $89\%$ & $66.7\%$ & $62\%$ & $\mathbf{71\%}$ & $\mathbf{80\%}$ & $67\%$ & $\mathbf{100\%}$ & $\mathbf{86.7\%}$ & $\mathbf{88\%}$ & $\mathbf{86\%}$ & $80\%$ & $75\%$ & $\mathbf{86.5\%}$\\
  HGNN \citep{feng:2019}    & $80\%$ & $50\%$ & $\mathbf{100\%}$ & $66.7\%$ & $75\%$ & $57\%$ & $\mathbf{80\%}$ & $78\%$ & $83\%$ & $\mathbf{86.7\%}$ & $\mathbf{88\%}$ & $\mathbf{86\%}$ & $78.3\%$ & $72.7\%$ & $81.5\%$\\
  \textbf{HUNet (ours)} & $80\%$ & $67\%$ & $89\%$ & $\mathbf{80\%}$ & $\mathbf{88\%}$ & $\mathbf{71\%}$ & $\mathbf{86.7\%}$ & $\mathbf{89\%}$ & $83\%$& $\mathbf{86.7\%}$ & $\mathbf{88\%}$ & $\mathbf{86\%}$  & $\mathbf{83.3\%}$ & $\mathbf{83\%}$ & $82.2\%$\\    
   \hline
\end{tabular}
\caption{ Classification results using different views from the morphological connectomic data. ACC: accuracy. SEN: sensitivity. SPEC: specificity. }\label{table:2} 
\end{table} 
\begin{table}[ht]
\centering
\begin{tabular}{cccc} 
\emph{Model} & \emph{Accuracy} & \emph{Sensitivity} & \emph{Specificity}  \\\midrule
GUNet \citep{gao:2019}   & $65\%$ & $65\%$ & $\mathbf{65\%}$ \\ 
 HGNN \citep{feng:2019} & $66\%$ & $\mathbf{90\%}$ & $43\%$ \\ 
 \textbf{HUNet (ours)} & $\mathbf{69\%}$ &  $86\%$ & $53\%$ \\\bottomrule

\end{tabular}
\caption{ Classification results using the ABIDE functional connectomic data.}\label{table:3} 
\end{table}

\squeezeup
\squeezeup

\section{Discussion}

We have presented HUNet, a hypergraph embedding architecture for learning high-order representative data embeddings that surpasses state-of-the art network embedding frameworks. We proposed our embedding architecture, designed to avoid the inability of existing deep graph embedding architectures to learn from the \emph{many-to-many} relationships between different nodes (i.e., data samples).  With our proposed framework, we treated the brain graph of each patient as a node in a hypergraph structure and learned a feature embedding which recapitulates the higher-order relationships prevalent between different subjects and used this feature embedding to classify nodes (i.e., subjects) into different brain states. Finally we demonstrated the outperformance of our method on two different types of brain connectomic datasets for neurological disorder diagnosis.

Our experimental results showed that HUNet was able to improve on GUNet architecture by an average of $3.3$\% on the small-scale morphological dataset, achieving up to a $13.3$\% difference in terms of classification accuracy in the individual views as shown in  (\textbf{Table.}~\ref{table:2}). We also demonstrated that HUNet outperforms GUNet on the large-scale functional network dataset by a margin of $4$\% as listed in (\textbf{Table.}~\ref{table:3}). This supports our claim of improving on graph-based methods through the use of hypergraphs to preserve high-order sample interactions when mapping data samples into a low-dimensional space. We also compared our method to HGNN architecture, which is the first geometric deep learning model on hypergraph structure. Clearly, HUNet outperformed the HGNN architecture \citep{feng:2019} by an average of $5$\% on the small-scale morphological dataset, demonstrating improvements of up to $13.3$\% in the individual views in terms of classification accuracy (\textbf{Table.}~\ref{table:2}). We also observe an improvement of $3$\% on the large-scale functional network dataset as shown in (\textbf{Table.}~\ref{table:3}). This further demonstrates that our hypergraph feature embedding architecture leveraging pooling and unpooling layers through the use of a U-Net encoder-decoder architecture is able to learn more representative and discriminative embeddings of the data features in comparison with solely relying on hypergraph convolutions for learning the hypernode embeddings as for HGNN \citep{feng:2019}.

\textbf{Limitations and recommendations for future work}\,\,
The parameters to optimize in the design of HUNet design architecture include the depth of the HUNet and the pooling ratios used in the hPool layers. Although we demonstrated the generalizability and scalability of our method, we note that if the depth parameter is increased too much the generalization ability weakens, which results in over-fitting. On the other hand, a large decrease in the pooling ratio, which means lowering the number of nodes to keep at each level of the HUNet, may result in having too few nodes to train on when coupled with a high depth parameter --particularly for small-scale datasets. In future work, one can integrate a \emph{hyper} attention mechanism to improve the quality of our hyper pooling and unpooling layers as in graph attention network \citep{velivckovic2017graph}.

\textbf{Broader impact}\,\, Dimensionality reduction or sample embedding is a fundamental step in many machine learning tasks such as classification, regression, and clustering to overcome the curse of dimensionality. Hence, learning how to embed samples into a low-dimensional space will have a broader impact in many real-world artificial intelligence applications. In this work, we leveraged the structure of a hypergraph to incorporate the high-order relationships existing among subsets of samples in real-life data. In fact, hypergraph representation learning  has been lagging behind compared to its graph counterpart in geometric deep learning  \citep{bronstein:2017}. In this work we extended the field of hypergraph representation learning to encoder-decoder architectures through the generalization of pooling and unpooling operations to hypergraphs within a U-Net architecture. 

Specifically, we addressed a fundamental scientific question: \emph{How to encode and decode many-to-many high-order relationships present in real life datasets to improve predictive learning tasks?} More importantly, we rooted the application of our proposed hypergraph encoder-decoder architecture in the field of network neuroscience with the aim of driving precision medicine forward. Our interdisciplinary work combined three research fields: data embedding, network neuroscience, and precision medicine, having different societal and economic impacts. \emph{First}, it propelled the development of \emph{automated} neurological disorder diagnosis systems. This can alleviate the societal burden of brain disorders by ensuring early and accurate automated diagnosis for effective treatment. Our proposed HUNet architecture treats the subjects as different nodes in a hypergraph, giving insights into the disordered brain alterations in neurological disorder patients, which can help tease apart variations in disorders. This will impact the future of \emph{disordered} brain connectivity related treatment and diagnosis methods. \emph{Second}, the development of the field of hypergraph-based data embedding can widen the horizon of geometric deep learning and its applications to complex samples with various interaction patterns.

\section{Conclusion}


In this paper we proposed the HUNet architecture, where the key idea is to learn a \emph{many-to-many} node embedding with a high-order feature aggregation rule by leveraging the structure of hypergraphs as they are able to model high-order interactions between subsets of nodes compared to existing deep graph-based embedding methods. With this mindset, we proposed the hypergraph pooling (hPool) and hypergraph unpooling (hUnpool) layers and generalized the U-Net architecture to hypergraphs. Using HUNet, we outperformed state-of-the-art graph and hypergraph sample embedding architectures using brain connectome datasets of varying scales, disorders, and distributions. Since our proposed HUNet architecture captures interactions between subsets of patients as hypernodes to pool in the feature embedding process, one can investigate the interpretability of the learned pooling and unpooling weights in order to group subjects with similar disordered brain alterations together. Designing an \emph{interpretable} HUNet as for generative adversarial networks (GANs) in \citep{Chen:2016} can help propel the field of precision medicine with the aim of gaining further insights into disordered brain alterations caused by neurological disorders and most importantly their variation across subsets of patients.  We refer interested readers to our GitHub HUNet source code available at \url{https://github.com/basiralab/HUNet}.

\section{Acknowledgements}

This work was funded by generous grants from the European H2020 Marie Sklodowska-Curie action (grant no. 101003403) to I.R.

\newpage
\bibliography{Biblio3.bib}
\bibliographystyle{model2-names}

\end{document}